\documentclass[12pt,a4paper]{article}
\usepackage[utf8]{inputenc}
\usepackage[english]{babel}
\usepackage{amsmath, amssymb, amsfonts}
\usepackage{amsthm} 
\usepackage{physics}
\usepackage[T1]{fontenc}
\usepackage{lmodern}
\usepackage[outline]{contour}
\usepackage{graphicx}
\usepackage{cite}
\usepackage[colorlinks=true, linkcolor=blue, citecolor=red, urlcolor=blue]{hyperref}
\usepackage{hyperref}
\usepackage{tikz}
\usepackage{geometry}
\usepackage{dsfont}
\usepackage{float}
\usepackage{cleveref}
\usepackage{authblk}

\theoremstyle{plain}

\theoremstyle{definition}

\theoremstyle{remark}

\author{Julio Cesar Jaramillo Quiceno}
\title{\textbf{$q$-Deformed Heisenberg Picture Equation}}
\begin{document}
\date{}
\maketitle

\begin{abstract}
In this paper we introduce the $q$-deformed Heisenberg picture equation.  We consider some examples such as : the spinless particle, the electrón interaction with a magnnetic field and $q$-deformed harmonnic oscillator.  The $q$-Heisenberg picture equation for any dynamical function at the end of the paper.
\end{abstract}

\vspace{0.6cm}

\section{Introduction}

The pictures encountered most usually in quantum mechanics are: the Schrödinger picture, and the Heisenberg picture. Let us consider the $q$- Hemitian Hamiltonian operator $\hat{H}_{q}$, which is depend explicitly on time, but operators do not, given by the {\em Schrödinger picture equation}

\begin{equation}
i\hbar\frac{\mathrm{d}}{\mathrm{d}t}\vert\psi_{q}(t)\rangle =\hat{H}_{q}\vert\psi_{q}(t)\rangle ,
\end{equation}

where $\vert\psi_{q}(t)\rangle$ denotes the $q$-state of the system in the Schrödinger picture, and can be expressed by means of the time-evolution operator, denoted  by $\hat{U}(t, t_{0})\vert \psi 
_{q}(t_{0})\rangle$, this is

\begin{equation}\label{U}
\hat{U}(t, t_{0}) = \exp\left[-i(t-t_{0})\hat{H}_{q}/\hbar\right],
\end{equation}

and satisfies some properties (see \cite{Zettili2009})

\begin{align}
\label{U1}\hat{U}(t, t_{0}) & = \ I,\\
\label{U2}\hat{U}^{\dagger}(t, t_{0}) & = \ \hat{U}(t_{0}, t).
\end{align}

For the Heisenberg picture,  is obtained from the Schrödinger picture by applying (\ref{U}) on $\vert\psi_{q}(t)\rangle_{H}$ 

\begin{equation}\label{Heis1}
\vert\psi_{q}(t)\rangle_{H} = \exp\left[it\hat{H}_{q}/\hbar\right]\vert\psi_{q}(t)\rangle ,
\end{equation}

thus, let us see how the expectation value of an operator $\hat{\mathcal{B}}$ in the state $\vert\psi_{q}(t)\rangle$ \cite{Zettili2009} and
consequently, one can introduce the Heisenberg picture as the average value of the time dependent physical observable $\hat{\mathcal{B}}$ 
\begin{equation}
\langle\psi_{q}(t)\vert\hat{\mathcal{B}}\vert\psi_{q}(t)\rangle = _{H}\langle\psi_{q}(t)\vert\hat{\mathcal{B}}^{H}(t)\vert\psi_{q}(t)\rangle_{H},
\end{equation}

where $\hat{\mathcal{B}}^{H}(t)$ is given by  \cite{Zettili2009}
\begin{equation}
\hat{\mathcal{B}}^{H}(t) = \hat{U}^{\dagger}\hat{\mathcal{B}}\hat{U}.
\end{equation}

The time evolution of $\mathcal{B}(t)$, i.e., the Heisenberg equation of motion or {Heisenberg picture equation} thus has the usual formulation
\begin{equation}\label{heisenberg-picture}
\frac{\mathrm{d}\hat{\mathcal{B}^{H}}}{\mathrm{d}t}  = \frac{1}{i\hbar}\left[\hat{\mathcal{B}^{H}}, \hat{H}_{q}\right].
\end{equation}

Note that the commutator $\left[\hat{\mathcal{B}^{H}}, \hat{H}_{q}\right]$ is not deformed, which means that these relation is subject to canonical time dependent  Heisenberg algebra, for instance to the coordinate and the momentum, $\hat{x}(t)$ and $\hat{p}(t)$ satisfy \cite{Miao-Xu2016} 
\begin{align}
\left[\hat{x}(t), \hat{p}(t)\right] & = \  i\hbar, \\
 \left[\hat{x}(t), \hat{x}(t)\right] &  = \  \left[\hat{p}(t), \hat{p}(t)\right] = 0.
\end{align}

The authors of the references \cite{Lavagno-Scarfone-Swamy2006, Wess-Madore1999} describes the $q$- deformed Heisenberg algebra for two generators $\tilde{x}$ and $\tilde{p}$  
\begin{align}
\left[\tilde{x}, \tilde{p}\right]_{q} & = \  i\tilde{\Lambda}_{q}, \\
\tilde{\Lambda}_{q}\tilde{x} & = \ q^{-1}\tilde{x}\tilde{\Lambda}_{q}, \\
\tilde{\Lambda}_{q}\tilde{p} & = \ q\tilde{p}\tilde{\Lambda}_{q}, \\
\left[\tilde{x}, \tilde{p}\right]_{q} & = \ q^{1/2}\tilde{x}\tilde{p}-q^{-1/2}\tilde{p}\tilde{x}.
\end{align}

This work was intended as an attempt to motivate the study of  the deformation of the Heisenberg picture equation, and is organized  of the following maner: In Section \ref{Heis-def1}, we show that the $q$-deformed Heisenberg picture equation, with their respective related formulas. While Section \ref{examples} some physical applications: spinless particle (subsection \ref{spin-part}), Electron interaction with a magnetic field (subsection \ref{electron}), and $q$-deformed harmonic oscillator (subsection \ref{harmonic}). Finally in the Section \ref{dynamical} also mentions the $q$-Heisenberg picture equation for any dynamical function.

\section{$q$-Deformed Heisenberg Picture Equation}\label{Heis-def1}
On the basis of the above properties, we have now the recipe  to generalize the definition of observables and to introduce a $q$-deformed Heisenberg picture equation in the framework of $q$-deformed theory.  Consistently with the standard quantum mechanics, there are many representations of wave functions and operators in quantum mechanics. The connection between the various representations is provided by unitary transformations. Each class of representation, also called a {\em picture}, differs from others in the way it treats the time evolution of the system \cite{Zettili2009}. This representation is  useful when describing phenomena with time dependent $q$-deformed Hamiltonians. Let us now derive the $q$-deformed equation of motion that regulates the time evolution of operators within the Heisenberg picture. Let $\hat{\mathcal{B}}_{q}$ be a $q$- Hermitian linear operator,  does not depend explicitly on time and since $\hat{U}_{q}(t)$ is unitary. Therefore the  $q$-deformed Heisenberg picture equation is given by

\begin{equation}\label{q-heisenberg-picture}
\frac{\mathrm{d}\hat{\mathcal{B}^{H}}}{\mathrm{d}t}  = \frac{1}{i\hbar}\left[\hat{\mathcal{B}^{H}}, q\hat{H}_{q}\right]_{q},
\end{equation}

where $\hat{\mathcal{B}^{H}}_{q} = \hat{U}^{\dagger}_{q}(t) \hat{\mathcal{B}}_{q}\hat{U}_{q}(t)$. This expression must be consistent to with the standard quantum mechanics.  On other hand, we postulate that the dynamics in the deformed phase space is described, in analogy with classical mechanics, by means of the following $q$-deformed evolution equations written in the form in virtue of (\ref{q-heisenberg-picture})

\begin{align}
\label{Hx}\frac{\mathrm{d}\hat{x}_{H}}{\mathrm{d}t} & = \ \frac{1}{i\hbar}\left[\hat{x}(t), q\hat{H}_{q}\right]_{q}, \\
\label{Hp}\frac{\mathrm{d}\hat{p}^{H}}{\mathrm{d}t} & = \ \frac{1}{i\hbar}\left[\hat{p}(t), q\hat{H}_{q}\right]_{q},
\end{align}
and the $q$-Heisenberg algebra for the operators $\hat{x}, \hat{p}$
\begin{align}
\label{qHx} q\hat{x}\hat{H}_{q}-q\hat{H}_{q}\hat{x} & = \ \left[\hat{x}(t), q\hat{H}_{q}\right]_{q},\\
\label{qHp} q\hat{p}\hat{H}_{q}-q\hat{H}_{q}\hat{p} & = \ \left[\hat{p}(t), q\hat{H}_{q}\right]_{q}.
\end{align}
Within this formalism let us discuss some examples.

\section{Some examples}\label{examples}

\subsection{Spinless particle}\label{spin-part}

As a ﬁrst simple example, we consider a spinless particle of mass $m$, which is moving in a one - dimensional inﬁnite potential well. For this case, Since the particle´s  Hamiltonian is purely kinetic, this is $\hat{H}_{q}=\frac{\hat{p}^{2}}{2m}$. We have $[\hat{p}(t), q\hat{H}_{q}]_{q}=0$ and using the relations for the $q$- Heisenberg algebra proposed in \cite{Wess-Madore1999, Bardek-Meljanac2000}, we get
\begin{equation*}
\left[\hat{x}(t), q\hat{H}_{q}\right]_{q} = \frac{q}{2m}\left[\hat{x}, \hat{p}^{2}\right]=\frac{i\hbar}{2m}q(q+1)\hat{p}\hat{\Lambda}_{q},
\end{equation*}
using (\ref{Hx}) and (\ref{Hp}) and integrating  we obtain 
\begin{equation}
\hat{x}_{H} = \hat{x}+ \frac{1}{2m}q(q+1)\hat{p}\hat{\Lambda}_{q}t, \quad \hat{p} = \hat{p}_{H}(t).
\end{equation}
\subsection{Electron interaction with a magnetic field}\label{electron}
 
As other example, we consider the the $q$-Hamiltonian due to the interaction of a electron of mass $m$, charge $e$, and spin $\hat{\textit{\textbf{S}}}$ with a magnetic ﬁeld pointing along the $z-$ axis is $\hat{H}_{q}=-(qeB/m_{e}c)\hat{S}_{z}$. The $q$-Heisenberg for  $\hat{H}$ with the components of the spin operator can be inferred at once from
\begin{eqnarray}
\label{SxSySz}\left[\hat{S}_{x}, \hat{S}_{y}\right]_{q} = -i\hbar \hat{\Lambda}_{q}\hat{S}_{z},\quad  \left[\hat{S}_{y}, \hat{S}_{z}\right]_{q} = i\hbar \hat{\Lambda}_{q}\hat{S}_{x}, \quad \  \left[\hat{S}_{x}, \hat{S}_{z}\right]_{q} = -i\hbar \hat{\Lambda}_{q}\hat{S}_{y},\\
\label{SH}\left[\hat{S}_{z}, \hat{H}_{q}\right]_{q} =0 , \quad \left[\hat{S}_{x}, q\hat{H}_{q}\right] = \frac{i\hbar eB}{m_{e}c}q^{2}\hat{\Lambda}_{q}\hat{S}_{y}, \quad \left[\hat{S}_{y}, q\hat{H}_{q}\right] = -\frac{i\hbar eB}{m_{e}c}q^{2}\hat{\Lambda}_{q}\hat{S}_{x},
\end{eqnarray}

using (\ref{q-heisenberg-picture}), leads to
\begin{equation}\label{sss}
\frac{\mathrm{d}\hat{S}_{x}}{\mathrm{d}t} = \frac{Bq^{2}\hat{\Lambda_{q}}}{m_{e}c}\hat{S}_{y}, \quad \frac{\mathrm{d}\hat{S}_{y}}{\mathrm{d}t}  =  - \frac{Bq^{2}\hat{\Lambda_{q}}}{m_{e}c}\hat{S}_{x},\quad  
\frac{\mathrm{d}\hat{S}_{z}}{\mathrm{d}t}  = 0,
\end{equation}

To solve (\ref{sss}), we may combine them into two more conducive equations

\begin{equation}\label{dst}
\frac{\mathrm{d}\hat{S}_{\pm}(t)}{\mathrm{d}t}=\pm i\frac{Bq^{2}\hat{\Lambda_{q}}}{m_{e}c}\hat{S}_{\pm}(t),
\end{equation}

where $\hat{S}_{\pm}(t)=\hat{S}_{x}(t)\pm i\hat{S}_{y}(t)$. Therefore, the solution of (\ref{dst}) leads to

\begin{align*}
\hat{S}_{x}(t) = \hat{S}_{x}(0)\cos (\omega_{q}t)-\hat{S}_{y}(0)\sin (\omega_{q}t),\\
 \hat{S}_{y}(t) = \hat{S}_{y}(0)\cos (\omega_{q}t)+\hat{S}_{0}(0)\sin (\omega_{q}t),\\
 \hat{S}_{z}(t) =0,
\end{align*}

being  $\omega_{q}=\frac{Bq^{2}\hat{\Lambda_{q}}}{m_{e}c}$.

\subsection{$q$-Deformed harmonic  oscillator}\label{harmonic}

The $q$-deformed harmonic oscillator  is described by the $q$-Hamiltonian

\begin{equation}\label{harmonicHam}
\hat{H}_{q}=\frac{\hbar \omega_{q}}{2}(\hat{a}\hat{a}^{\dagger}+\hat{a}^{\dagger}\hat{a}),
\end{equation}

where $\omega_{q}=\omega [2]_{q}/2q^{2}$ is the  $q-$ frequency harmonic oscillator \cite{Lavagno-Scarfone-Swamy2006}, and $\hat{a}, \hat{a}^{\dagger}$ are creation and annihilation operators satisfying the $q-$ commutation relation   $\hat{a}\hat{a}^{\dagger}-q\hat{a}^{\dagger}\hat{a}=1$. Using (\ref{q-heisenberg-picture}) we have

\begin{equation}\label{a+a}
\frac{\mathrm{d}\hat{a}}{\mathrm{d}t}=-iq\omega_{q}\hat{a}, \quad \ \frac{\mathrm{d}\hat{a}^{\dagger}}{\mathrm{d}t}=iq\omega_{q}\hat{a}^{\dagger},
\end{equation}

therefore, the solutions of (\ref{a+a}) is given by

\begin{equation}
\hat{a}_{H}(t) =\hat{a}(0)\exp [-iq\omega_{q}t], \quad \hat{a}^{\dagger}_{H}(t) =\hat{a}(0)\exp [iq\omega_{q}t].
\end{equation}

This representation satisfies the $q$-commutation relation $\hat{a}\hat{a}^{\dagger}-q\hat{a}^{\dagger}\hat{a}=1$, and yields the standar harmonic oscillator expressions in the limit as $q\longrightarrow 1$. On other hand taking into account the  coordinate and momentum operators in $q$-deformed quantum mechanics, we have in the Heisenberg picture

\begin{align}
\hat{x}_{H}(t) & = \ \sqrt{\frac{\hbar}{2m\omega_{q}}}(\hat{a}_{H}(t) +\hat{a}^{\dagger}_{H}(t) )=2\sqrt{\frac{\hbar}{2m\omega_{q}}}\hat{a}(0)\cos (q\omega_{q}t), \\
\hat{p}_{H}(t) & = \ i\sqrt{\frac{m\omega_{q}\hbar}{2}}(\hat{a}_{H}(t)- \hat{a}^{\dagger}_{H}(t))= 2\sqrt{\frac{m\omega_{q}\hbar}{2}}\hat{a}^{\dagger}(0)\sin (q\omega_{q}t).
\end{align}

In the following  section, we introduce the $q-$ Heisenberg picture equation for any dynamical function.

\section{$q$-Heisenberg  picture equation for any dynamical function}\label{dynamical}

Let $\hat{x}, \hat{y}, \tilde{x}, \tilde{y}$ be the generators, and $u$ an observable  of $q$-Heisenberg algebra

\begin{align}
\label{h1}q^{1/2}\hat{x}\hat{y}-q^{-1/2}\hat{y}\hat{x} & = \ i\Lambda_{q},\\
\label{h2}q^{1/2}\tilde{x}\tilde{y}-q^{-1/2}\tilde{y}\tilde{x} & = \ i\tilde{\Lambda}_{q},\\
\label{h3}\hat{x}\tilde{x} & = \ \tilde{x}\hat{x},\\
\label{h4}\hat{y}\tilde{y} & = \ \tilde{y}\hat{y},
\end{align}

for  any dynamical function which is defined as $\hat{f}(\hat{x}(u),\hat{y}(u), \tilde{x}(u), \tilde{y}(u))$, can be described by means of the following evolution equations or {\em q- Heisenberg picture equations}

\begin{equation}\label{dyn}
\frac{\mathrm{d}\hat{f}}{\mathrm{d}u} = q\hat{f}\hat{H}_{q}-q\hat{H}_{q}\hat{f},
\end{equation}

following \cite{Lavagno-Scarfone-Swamy2006}, in fact, we recall readily that the most general form of a function $\hat{f}\in\mathcal{P}$ can be written as a polynomial in the $q$- variables $\hat{x}(u), \hat{y}(u), \tilde{x}(u)$ and $\tilde{y}(u)$

\begin{align}
\label{alfa}\hat{f}(\hat{x}(u), \hat{y}(u); u) & = \ \sum\limits_{n,m}\alpha_{mn}(u)\left[\hat{x}(u)\right]^{n}\left[\hat{y}(u)\right]^{m},\\
\label{beta}\hat{f}(\tilde{x}(u), \tilde{y}(u); u) & = \ \sum\limits_{n,m}\beta_{mn}(u)\left[\tilde{x}(u)\right]^{n}\left[\tilde{y}(u)\right]^{m}.
\end{align}

where $\alpha_{mn}(u)$ and $\beta_{mn}(u)$ are $C$-numbers which may be $u-$ dependent and we have assumed the  $\hat{x}, \hat{y}$ ordering prescription which can be always accomplished by means of Eq. (\ref{h1}). Its $u$-derivative becomes

\begin{equation}\label{Dxy}
\dfrac{\mathrm{d}}{\mathrm{d}u}\left(\sum\limits_{n,m}\alpha_{mn}(u)\left[\hat{x}(u)\right]^{n}\left[\hat{y}(u)\right]^{m} \right) = \left[\sum\limits_{n,m}\alpha_{mn}(u)\left[x(u)\right]^{n}\left[y(u)\right]^{m}, q\hat{H}_{q}\right]
\end{equation}
\begin{equation}\label{Dxytilde}
\dfrac{\mathrm{d}}{\mathrm{d}u}\left(\sum\limits_{n,m}\beta_{mn}(u)\left[\tilde{x}(u)\right]^{n}\left[\tilde{y}(u)\right]^{m}\right) = 
\left[\sum\limits_{n,m}\beta_{mn}(u)\left[\tilde{x}(u)\right]^{n}\left[\tilde{y}(u)\right]^{m}, q\hat{H}_{q}\right].
\end{equation}

Now, as example of application,  we consider an  arbitrary system described by the $q$-Hamiltonian of the form  $\hat{H}_{q}=b\hat{x}+c\hat{y}, b,c\in\mathbb{R}$.   For $n=1,m=0$ we have 

\begin{align}\label{Dxy10}
\begin{split}
\dfrac{\mathrm{d}}{\mathrm{d}u}\left(\alpha_{10}(u)\hat{x}(u)\right)  & = \  \left[\alpha_{10}(u)\hat{x}(u), q(b\hat{x}(u)+c\hat{y}(u)) \right]_{q},\\
 & = \  ic\alpha_{10}(u)\Lambda_{q}q^{1/2},
\end{split}
\end{align}

solving (\ref{Dxy10}) for $u=0$ and $u=t$ we obtain

\begin{equation}
\alpha_{10}(t)\hat{x}(t)=\alpha_{10}(0)\hat{x}(0)\exp\left[i\Lambda_{q}q^{1/2}c\int^{t}_{0}\alpha_{10}(u)\mathrm{d}u\right].
\end{equation}

Now, for $n=0, m=1$ the differential equation is given by

\begin{align}\label{Dxy01}
\begin{split}
\dfrac{\mathrm{d}}{\mathrm{d}u}\left(\alpha_{01}(u)\hat{y}(u)\right)  & = \  \left[\alpha_{01}(u)\hat{y}(u), q(b\hat{x}(u)+c\hat{y}(u)) \right]_{q},\\
 & = \ -ibq^{3/2}\Lambda_{q}\alpha_{01}(u)
\end{split}
\end{align}

the same way as the above case,  their respective solution can be written as

\begin{equation}
\alpha_{01}(t)\hat{y}(t)=\alpha_{01}(0)\hat{y}(0)\exp\left[-i\Lambda_{q}q^{3/2}b\int^{t}_{0}\alpha_{10}(u)\mathrm{d}u\right].
\end{equation}

and finally for the general case we have 

\begin{align}\label{sol}
\begin{split}
\sum\limits_{n,m}\alpha_{nm}(t)x^{n}(t)y^{m}(t) & = \\
\sum\limits_{n,m}\alpha_{nm}(0)x^{n}(0)y^{m}(0)+i\Lambda_{q}\int_{0}^{t}\sum\limits_{n,m}\alpha_{n,m}(u)\left[q^{3/2}c[n]_{q}\hat{x}^{n-1}(u)-q^{1/2}[m]_{q}b\hat{y}^{m-1}(u)\right]\mathrm{d}u.
\end{split}
\end{align}

where Eq. (\ref{h1}) has been employed and where we have introduced the $q$- basic numbers

\begin{equation}
[n]_{q} = \frac{q^{2n}-1}{q^{2}-1}, \quad [m]_{q} = \frac{q^{2m}-1}{q^{2}-1}.
\end{equation}

On other hand , the general solutions for (\ref{Dxy}) and (\ref{Dxytilde}) inn terms of the $q$-Hamiltonian can be expressed as

\begin{align}
\sum\limits_{n,m}\alpha_{nm}(t)\hat{x}^{n}(t)\hat{y}^{m}(t) & = \ \sum\limits_{n,m}\alpha_{n,m}(0)\hat{x}^{n}(0)\hat{y}^{m}(0)\exp\left[q\int_{0}^{t}\sum\limits_{n,m}\alpha_{nm}(u)\frac{1}{\hat{x}^{n}\hat{y}^{m}}\left[\hat{x}^{n}\hat{y}^{m}, q\hat{H}_{q}\right]_{q}\mathrm{d}u\right],\\
\sum\limits_{n,m}\beta_{nm}(t)\tilde{x}^{n}(t)\tilde{y}^{m}(t) & = \ \sum\limits_{n,m}\beta_{n,m}(0)\tilde{x}^{n}(0)\tilde{y}^{m}(0)\exp\left[q\int_{0}^{t}\sum\limits_{n,m}\beta_{nm}(u)\frac{1}{\tilde{x}^{n}\tilde{y}^{m}}\left[\tilde{x}^{n}\tilde{y}^{m}, q\hat{H}_{q}\right]_{q}\mathrm{d}u\right].
\end{align}

Since $q-$ Heisenberg’s equations is first order, its formal solution contains two  constant operator namely $\sum\limits_{n,m}\alpha_{n,m}(0)\hat{x}^{n}(0)\hat{y}^{m}(0)$ and $\sum\limits_{n,m}\beta_{n,m}(0)\tilde{x}^{n}(0)$.  Note that the structure of the $q$-Heisenberg equations (\ref{Dxy}) and (\ref{Dxytilde}) are similar to the classical equation of motion of a dynamical function $\hat{f}$ that does not depend explicitly on $u-$ variable  $\mathrm{d}\hat{f}/\mathrm{d}u=\left\lbrace \hat{f}, \hat{H}_{q}\right\rbrace$  , where $\left\lbrace \hat{f}, \hat{H}_{q}\right\rbrace$ is a $q$-Poisson Bracket  between $\hat{f}$ and $\hat{H}_{q}$ \cite{Lavagno-Scarfone-Swamy2006}. On other hand, In the Heisenberg picture the $u$- dependence is transferred from the wave functions to the operators with the usual $u$-dependence, therefore (\ref{dyn}) can be expressed in the form

\begin{equation}\label{dyn1}
-i\frac{\mathrm{d}\hat{f}}{\mathrm{d}u} = \left[q\hat{H}_{q}, \hat{f}\right]_{q},
\end{equation}

Since the above commutation of $\hat{H}_{q}$ with $\hat{f}$ is a linear transformation of the operator dynamical function $\hat{f}$, it may be written in terms of a $q-$ superoperator, $\mathcal{L}_{q}$, called the {\em Liouvillian for the dynamical function operator } $\hat{f}$ (see \cite{Lowdin1985} for the case non-deformed)

\begin{equation}\label{dyn2}
\mathcal{L}_{q}\hat{f} \equiv \left[q\hat{H}_{q}, \hat{f}\right]_{q},
\end{equation}  

The $q$-Liouvillian is a superoperator since it is a linear transformation acting on the space of observables or operators rather than on the space of wave functions. If the $q-$Hamiltonian is $u$- independent, then the $q$-Liouvillian is also $u$- independent and $q$-Heisenberg picture equation (\ref{dyn1}) has the formal solution

\begin{equation}
\hat{f}(u)=\exp\left[i\mathcal{L}_{q}u\right]\hat{f}(0).
\end{equation}

\bibliographystyle{plain} 
\bibliography{myBiblib}
\end{document}